\documentclass[structabstract]{aa}  
\usepackage{graphicx}
\usepackage{txfonts}
\usepackage{textcomp}
\usepackage{natbib}
\begin{document}
   \title{Star formation with disc accretion and rotation}
 \subtitle{I. Stars between 2 and 22 M$_\odot$ at solar metallicity}

     \author{L. Haemmerl\'e,      
     P. Eggenberger,
         G. Meynet, 
          A. Maeder,      
        C. Charbonnel
}

  \authorrunning{Haemmerl\'e et al.}

   \institute{Geneva Observatory, University of Geneva, Maillettes 51, CH-1290 Sauverny, Switzerland
  }

   \date{Received ; accepted }

 
 \abstract
{The way angular momentum is built up in stars during their formation process may have an impact on their further evolution.}
{In the frame of the cold disc accretion scenario,
we study for the first time how angular momentum builds up inside the star during its formation and what are
the consequences for its evolution on the main sequence (MS).}
{Computation begins from a hydrostatic core on the Hayashi line of 0.7 M$_\odot$ at solar metallicity (Z=0.014) rotating as a solid body.
Accretion rates depending on the luminosity of the accreting object are considered
varying between 1.5$\times$10$^{-5}$ and 1.7$\times$10$^{-3}\,$M$_\odot\,$yr$^{-1}$.
The accreted matter is assumed to have an angular velocity equal to that of the outer layer of the accreting star. Models are computed for a mass-range on the zero-age main sequence (ZAMS) between 2 and 22\,M$_\odot$.}
{We study how the internal and surface velocities vary as a function of time during the accretion phase and the evolution towards the ZAMS.
Stellar models, whose evolution has been followed along the pre-MS phase, are found to exhibit a shallow gradient of angular velocity on the ZAMS.
Typically, the 6 M$_\odot$ model has a core that rotates 50\% faster than the surface on the ZAMS.
The degree of differential rotation on the ZAMS decreases when the mass increases (for a fixed value of $\rm v_{\rm ZAMS}/v_{\rm crit}$).
The MS evolution of our models with a pre-MS accreting phase show no significant differences
with respect to those of corresponding models computed from the ZAMS with an initial solid-body rotation. 
Interestingly, for masses on the ZAMS larger than 8 M$_\odot$,
there exists a maximum surface velocity that can be reached through the present scenario of formation.
Typically, for 14 M$_\odot$ models, only stars with surface velocities on the ZAMS lower than about 45\% of the critical velocity can be formed.
To reach higher velocities would require to start from cores rotating above the critical limit.
We find that this upper velocity limit is smaller for higher masses.
In contrast, below 8 M$_\odot$, there is no restriction and the whole domain of velocities, up to the critical one, can be reached.}
 {}
 
 
\maketitle
%

\section{Introduction}
\label{sec-intro}
Stars are formed from the fragmentation and the collapse of large molecular clouds.
The collapse being non-homologous (\citealt{mcnally1964}, \citealt{bodenheimer1968}, \citealt{larson1969}),
hydrostatic equilibrium is first reached by a central core while the rest of the cloud is still collapsing.
A part of the remaining mass is then accreted by the central core while it is already evolving through a slow Kelvin-Helmholtz (KH) contraction,
the rest being ejected to infinity.
This is the star formation scenario through accretion
(\citealt{shu1977}, \citealt{stahler1980i}, \citealt{stahler1980ii}, \citealt{stahler1986i}, \citealt{stahler1986ii}).

It is currently believed that this scenario holds for low-, intermediate-mass stars and also
for massive stars (e.g. \citealt{nakano1989}, \citealt{bernasconi1996}, \citealt{yorke2008}, \citealt{kuiper2010}).
While the general picture seems to be well defined, our knowledge of the accretion history is far from being complete
and many questions regarding as how the accretion history depends on the physics of the cloud (metallicity, turbulence),
or of the already formed stars are still quite open.
Also the complex hydrodynamics of the accretion process itself is a very active area of research
(e.g. \citealt{peters2010}, \citealt{peters2011}, \citealt{kuiper2010}, \citealt{kuiper2011}, \citealt{girichidis2011}).

The aim of the present work is to study, for the first time in the frame of the accretion scenario,
the role of rotation during the pre-MS phase of intermediate-mass stars.
We would also like to know how the angular momentum builds up inside the forming star
and which kind of rotation profile is obtained when the star enters the core hydrogen-burning phase on the ZAMS,
defined as the point when central hydrogen abundance has decreased by 3\textperthousand.
One can indeed expect that depending on the accretion history,
different distributions of the angular velocity inside the star can be obtained with some impact on the evolution during the MS phase.

This work is the first one of a series of papers exploring these effects.
Here we consider the case of stars with masses on the ZAMS between 2 and 22 M$_\odot$ at solar metallicity.
In Sect. \ref{sec-Mod}, we describe the physical ingredients used for the computation
of our models, while the results are presented in Sect. \ref{sec-Res}. Implications of these results
for explaining the observed surface velocities of Herbig Ae/Be objects are discussed
in Sect. \ref{sec-HAeBe}. Possible impact of the pre-MS evolution on the evolution of stars
on the MS phase are presented in Sect. \ref{sec-MS} and conclusions are given in Sect. \ref{sec-Outro}.

\section{Physical ingredients of the models}
\label{sec-Mod}

The models are computed with the Geneva stellar evolution code (\citealt{eggenberger2008}),
with the same treatment as \cite{ekstroem2012} for the rotational instabilities,
and with the following modifications to account for the pre-MS phase:
\begin{itemize}
\item The burning of D, $^6$Li and $^7$Li is included,
through the following reactions:
\begin{equation}
\rm D\,(p,\gamma)\,^3He\quad D\,(D,p)\,^3H\,(\,,e^-\nu)\,^3He\quad D\,(D,n)\,^3He
\end{equation}
\begin{equation}
\rm^6Li\,(p\,,\,^3He)\,^4He
\end{equation}
\begin{equation}
\rm^7Li\,(p,\gamma)\,^8Be\quad^7Li\,(p,\alpha)\,^4He
\end{equation}
The corresponding reaction rates are those of \cite{caughlan1988}.
Let us recall that 
$^6$Li- and $^7$Li-burning have a negligible impact in terms of energy generation.
The main contribution to nuclear energy generation rate comes from D-burning.
This contribution becomes significant for temperatures in stellar interiors around 1-2$\times$10$^6$ K.

\item Mixing in convective zones is assumed to be instantaneous for all chemical species, except for D;
for this fragile element (see above), mixing is treated as a diffusive process
with a diffusion coefficient deduced from the mixing-length theory (\citealt{bernasconi1997}).
It is given by
\begin{equation}
\rm D_{MLT}=\frac{1}{3}\alpha^{4/3}H_P\left(\frac{9}{32}\frac{cg}{\rho\kappa}(1-\beta)\nabla_{ad}(\nabla_{rad}-\nabla_{ad})\right)^{1/3}
\end{equation}
where $\alpha$ is the mixing-length coefficient, H$\rm_P$ the pressure-scale height,
c the speed of light, g the gravitational acceleration,
$\rho$ the mass density, $\kappa$ the opacity, $\beta$ the ratio of gas pressure to total pressure,
$\nabla_{\rm ad}$ the adiabatic gradient and $\nabla_{\rm rad}$ the radiative gradient.
\item We suppose that accretion occurs through the scenario of cold disc accretion (\citealt{hosokawa2010}).
It means that matter is gently deposited at the surface of the star
with an internal energy content equal to that contained in the outer layers of the accreting stars.
This procedure is the same as in \cite{yorke2008} and \cite{hosokawa2010}.
It could result for instance from interactions between the stellar radiation field and the infalling mater,
leading to a thermal coupling between this material and the stellar photosphere.
The angular velocity $\Omega$ of the new material is also taken equal to that of the surface of the accreting star before the new layer is accreted.
This accretion law corresponds to the smoothest way to accrete angular momentum.
A weak magnetic field could be sufficient to drag the infalling matter along the field lines and to obtain such an accretion law for the angular momentum.\footnote{This assumption corresponds to the lower limit for the accretion of angular momentum, the upper limit being given by the case where the material is accreted with a quasi-keplerian velocity (i.e. quasi-critical velocity), which is expected to be the velocity of the material in the disc. We shall investigate the impact of this other limiting case in a forthcoming work.}
It turns out that the resulting time derivative of the total angular momentum, $\dot J$, is nearly proportional to the mass accretion rate, $\dot M$.
This implies that the mean specific angular momentum accreted at each evolution step is almost constant all along the birthline.
\end{itemize}

\subsection{Mass accretion rate}
\label{sec-dm}

We use the same accretion law as in \cite{behrend2001}.
Let us recall a few points about this accretion law. First, it is based on an empirical relation
established by \cite{churchwell1998} and confirmed by \cite{henning2000}, between the mass outflow
rate, $\dot M_{\rm out}$, and the bolometric luminosity, $L$, of the accreting object in ultra-compact HII regions:
\begin{equation}
\log\dot M_{\rm out}=-5.28+\log\frac{L}{\rm L_\odot}\cdot(0.752-0.0278\cdot\log\frac{L}{\rm L_\odot})
\label{eq-dm1}
\end{equation}
where $\dot M_{\rm out}$ is in solar masses per year.
To deduce an accretion rate from $\dot M_{\rm out}$, \cite{behrend2001} made the hypothesis that
a constant fraction $f$ of the mass collapsing from the accretion reservoir is accreted by the star.
The other fraction, $(1-f)$, produces the outflow of equation \ref{eq-dm1}.
Thus, we have a very simple relation between the rate of mass outflow and the rate of mass accretion
\begin{equation}
\dot M_{\rm acc}={f\over1-f}\dot M_{\rm out}.
\label{eq-dm2}
\end{equation}
In the following we adopt a value of 1/3 for $f$, as in \cite{behrend2001}.
This implies that the accretion rate is equal to half of the mass outflow.
Such an accretion law gives accretion rates between 1.5$\times$10$^{-5}$ M$_\odot$ yr$^{-1}$ for objects with luminosities around 10 L$_\odot$
and 10$^{-3}$ M$_\odot$ yr$^{-1}$ for luminosities equal to 10$^4$~L$_\odot$.

It has to be noted that the empirical relation obtained by \cite{churchwell1998}, and used here, is not necessarily describing an evolutionary sequence.
It may for instance result from a relation between the mass outflow rate and the luminosity
when the outflow rate reaches some maximum value and is thus detectable.
Implicitely we assume here that most of the star is built up during events following the above relation
and that other accretion phases (if any) would have a negligible impact.
Let us note that there are other arguments, observational and theoretical, for justifying the use of the present accretion law.
1) There are still many uncertainties on the accretion rates and we think that the present choice remains an interesting possibility
since the birthline resulting from such a law well fits the observed upper envelope in the HR diagram of the positions of Herbig Ae/Be objects
(see Fig. 1 in \citealt{behrend2001}). 
2) The present accretion rates are in the range of values expected by theoretical considerations.
The typical accretion rate $\dot M\sim M_{\rm J}/t_{\rm ff}\sim c_{\rm s}^3/G$ for a cloud temperature of 10-20 K is $\rm\sim10^{-5}\,M_\odot\,yr^{-1}$.
Such a rate is expected for solar-type stars.
For stars around $\rm50\,M_\odot$, the Kelvin-Helmholtz time is $\rm\sim5\times10^4\,yr$,
so that a simple argument about timescales ($M/\dot M<t_{\rm KH}$) imposes an accretion rate higher than $\rm\sim10^{-3}\,M_\odot/yr$.
3) Accretion rates in the same ranges have been used by recent similar works modeling massive star formation through accretion.
For instance \cite{hosokawa2009} calculate their pre-MS evolution with constant mass accretion rates
spanning values between 10$^{-6}$ and $6\times10^{-3}$ M$_\odot$ yr$^{-1}$.

\subsection{Initial model}
\label{sec-ini}

We begin the computations with a hydrostatic core of 0.7 M$_\odot$, as in \cite{behrend2001},
with $L$ = 7.69 L$_\odot$, $T_{\rm eff}$ = 3852 [K] and $Z$ = 0.014.
We take the chemical abundances of \cite{asplund2005}, and the Ne value from \cite{cunha2006}.
For the light elements, we take mass fractions of $X_2$ = 5$\times$10$^{-5}$, $X_6$ = 9$\times$10$^{-10}$ and $X_7$ = 1$\times$10$^{-8}$
($X_i$ being the mass fraction of the isotope of mass number $i$). This value for $X_2$ is the same as in \cite{behrend2001}.
The chemical composition of the accreted material is taken equal to the initial stellar composition.

This choice of $L$ and $T_{\rm eff}$ corresponds to the case of spherical accretion.
When the pre-stellar cloud collapses, the disc is expected to form only after an initial phase of spherical accretion,
so that the relevant scenario for the formation of the initial model is the scenario of spherical accretion,
even if we consider cold disc accretion during the rest of the pre-MS evolution.
We assume that a spherical accretion was responsible for the growing up of the star until a mass of about 0.7 M$_\odot$ is reached.
From this stage on, we pursue the evolution assuming that accretion proceeds through a cold disc.
In Sect. \ref{sec-vmax}, we briefly mention how the choice of an initial model built by disc accretion would modify our results,
but a more detailed study of the impact of the initial model is postponed to a forthcoming work.

The initial model is fully convective, and solid-body rotation is assumed in convective zones.
The rotation of the initial model is thus determined by a single angular velocity $\rm\Omega_{ini}$ for all layers,
that we can choose as a free parameter.
Since the accretion of angular momentum during the subsequent evolution is linked to the rotation of the accreting object,
the value we choose for $\rm\Omega_{ini}$ has an impact on the angular momentum accretion rate for all the subsequent phases.

\subsection{Models}
\label{sec-mod}

We compute models for different final masses on the ZAMS $M\rm_{ZAMS}$ from 2 M$_\odot$ to 22 M$_\odot$.
In each case, we begin with the initial model described in Sect. \ref{sec-ini}, that is accreting mass at a rate given by equations \ref{eq-dm1} and \ref{eq-dm2}.
When the mass of the star reaches the value $M\rm_{ZAMS}$,
we turn off accretion and the star continues its evolution at constant mass, while the star keeps contracting toward the ZAMS.
For each $M\rm_{ZAMS}$, we compute models with different $\rm\Omega_{ini}$, between 0 and $\rm2.25\times10^{-5}\,s^{-1}$.
This last value corresponds to the critical angular velocity of the initial hydrostatic core.\footnote{
   The critical angular velocity is defined as the angular velocity such that the centrifugal force balances the gravity:
   $\Omega_{\rm crit}=(GM/R^3_{\rm eq,crit})^{1/2}$.
   The critical velocity is defined as v$_{\rm crit}=R_{\rm eq,crit}\Omega_{\rm crit}$.
   }

\section{Results}
\label{sec-Res}

\subsection{The accretion phase}
\label{sec-bl}

\begin{figure}
\includegraphics[width=0.49\textwidth]{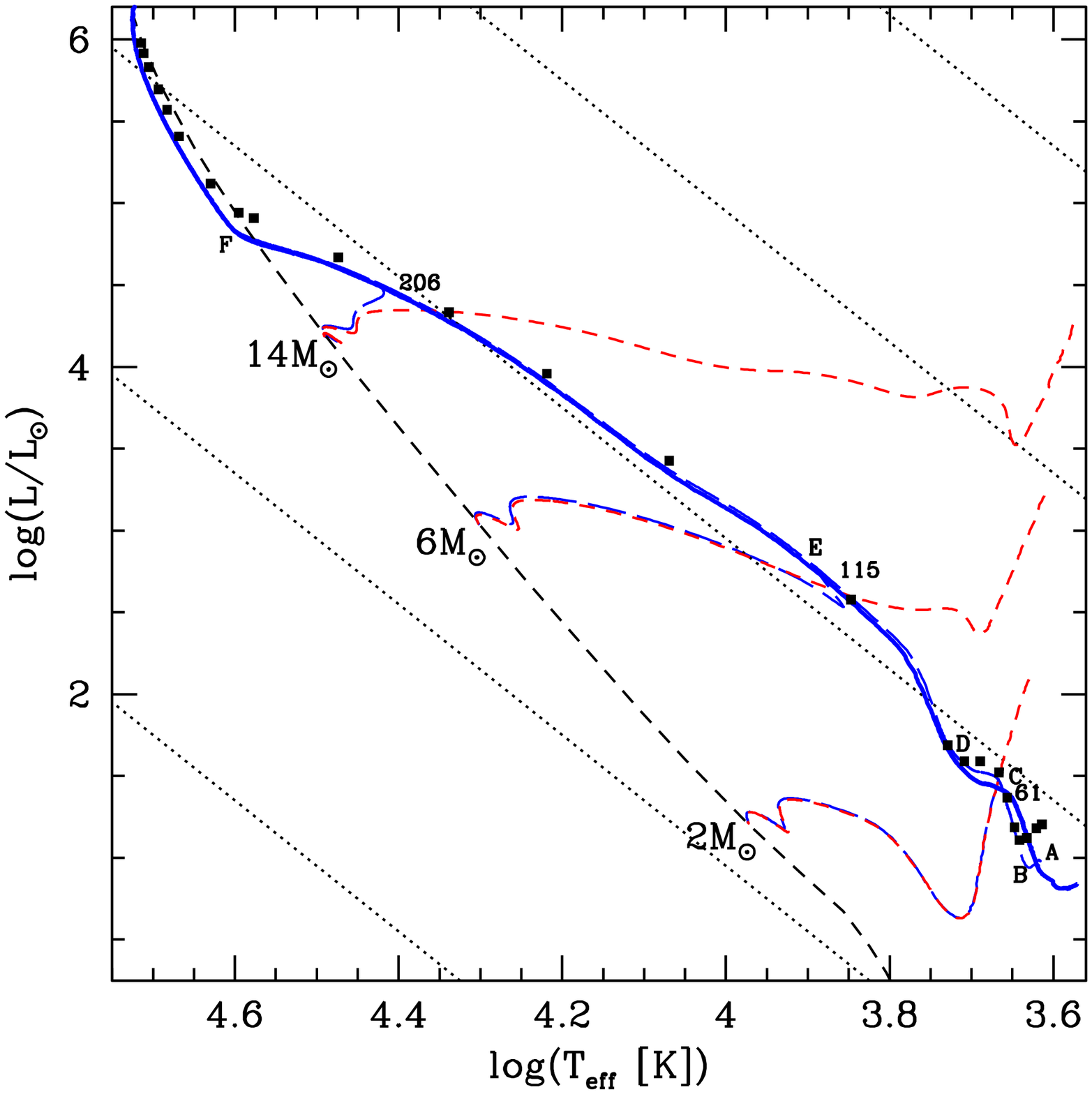}
\caption{
Birthline with rotation ($\rm\Omega_{ini}=2.1\times10^{-5}\,s^{-1}$, solid blue line),
and birthline without rotation, with three contractions (long-dashed blue lines).
Some points of the birthline of \cite{behrend2001} are displayed for comparison (filled squares).
The short-dashed lines indicate the ZAMS of \cite{ekstroem2012} without rotation (black),
and the pre-MS tracks at constant mass corresponding to the contractions displayed (red).
The numbers along the birthline are rotational velocities (in km s$^{-1}$) along the birthline with rotation
at $M$ = 2, 6 and 14 M$_\odot$.
Letters along the birthline indicate the evolutionary stages described in the text (Sect. \ref{sec-bl}).
Straight dotted lines are iso-radius of 0.1, 1, 10, 100 and 1000 R$_\odot$ from bottom left to top right.
}
\label{fig-hrbl}
\end{figure}

\begin{figure}
\includegraphics[width=0.49\textwidth]{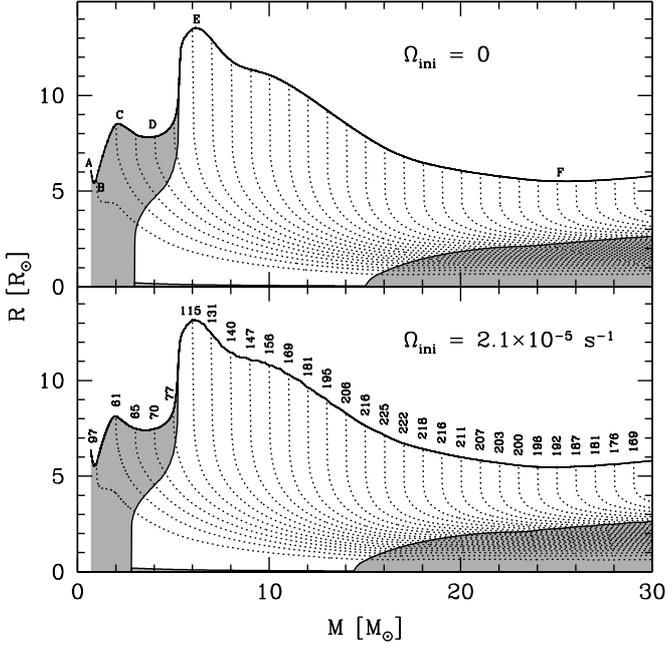}
\caption{
Evolution of the radius and the internal structure of the star on the birthline without rotation (upper panel)
and with rotation ($\rm\Omega_{ini}$ = 2.1$\times$10$^{-5}$ s$^{-1}$, lower panel).
In each panel, the upper solid line is the radius and shaded areas correspond to convective regions.
The dotted lines indicate the iso-masses, i.e. the radius corresponding to fixed values of the lagrangian coordinate $M\rm_r$ ($M\rm_r$ = 1, 2, 3, ... [M$_\odot$]).
Letters along the radius on the upper panel indicate the evolutionary stages described in the text (Sect. \ref{sec-bl}).
Numbers along the radius on the lower panel correspond to rotational velocities (in km s$^{-1}$) at $M$ = 1, 2, 3, ... [M$_\odot$].
}
\label{fig-stbl}
\end{figure}

The birthline without rotation and the birthline with $\rm\Omega_{ini}=2.1\times10^{-5}\,s^{-1}$ are visible on Fig. \ref{fig-hrbl}.
Note that the shape of the birthline as well as the sequence of events described below
would be very similar for other choices of the core initial angular velocity.
For comparison, we added some points of the birthline of \cite{behrend2001}.
The evolution of the radius of the star and of its internal structure is given on Fig. \ref{fig-stbl}.
The evolution on the birthline follows roughly five stages, delimited by capital letters on Figs. \ref{fig-hrbl} and \ref{fig-stbl},
each one corresponding to a contraction or an expansion phase:
\begin{enumerate}
\item$A\rightarrow B:$
At the beginning of the evolution, the central temperature of the star is too low for nuclear reactions to occur at a significant rate.
Weak D-burning takes place in deep regions, but the corresponding energy generation rate $\rm\epsilon_D$ is negligible ($\sim10^{-16}$ [erg g$^{-1}$ s$^{-1}$]).
The energy comes from gravitational contraction, and luminosity decreases, as in the case of a classical Hayashi line.
\item$B\rightarrow C:$
When the central temperature reaches $\sim$ 10$^6$ [K], D-burning becomes significant and its abundance decreases in the centre.
The star expands and its luminosity increases.
\item$C\rightarrow D:$
When the central mass fraction of D becomes too low ($\lesssim$ 5$\times$10$^{-6}$),
the energy liberated by D-burning is no longer sufficient to sustain the star and gravitational contraction starts again.
The stellar luminosity stops growing for a while, but the central temperature continues to increase.
When the central temperature reaches $\sim$ 2.8$\times$10$^6$ [K], the opacity is low enough for radiative equilibrium to establish in the center.
\item$D\rightarrow E:$
While the radiative core grows, the changes in the structure of the star produce a steep increase of its radius.
This expansion is enhanced by the fact that the accreted D can no longer reach the center and burns now in an off-center shell.
\item$E\rightarrow F:$
Once the star is completely radiative, the radius decreases again, while its luminosity continues to grow.
A convective core develops before the ZAMS, when the energy liberated by the CN-cycle becomes significant.
\end{enumerate}

As we can see in Fig. \ref{fig-hrbl}, the differences between our birthline and that of \cite{behrend2001} are very modest
and are mainly due to differences in the initial chemical composition.
In \cite{behrend2001}, an initial metallicity of $Z$=0.020 was considered, while here we take $Z=0.014$.

\begin{figure}
\includegraphics[width=0.49\textwidth]{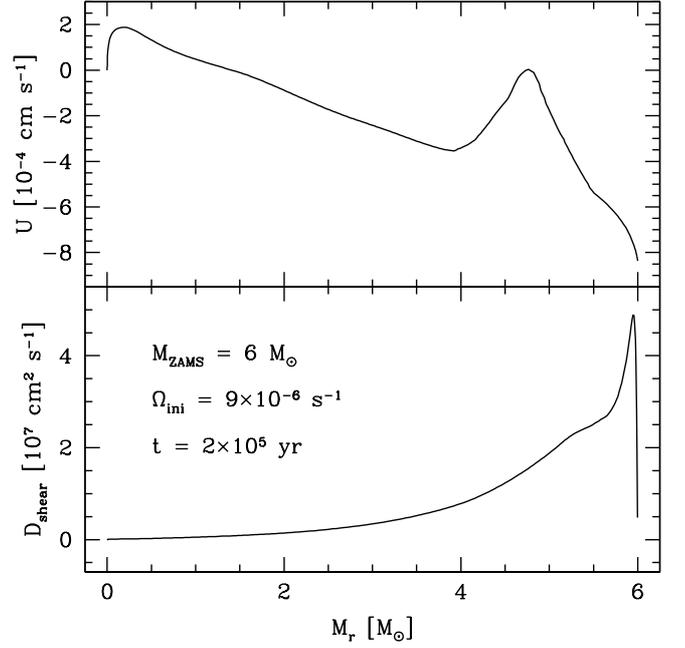}
\caption{
\textit{Upper panel:} Vertical component of the meridional circulation for the 6 M$_\odot$ model
with $\rm\Omega_{ini}=9\times10^{-6}\,s^{-1}$ (i.e. model of the grid described in Sect. \ref{sec-gr}),
at an age of 200 000 years, just at the end of the accretion phase (see Fig. \ref{fig-st6}).
\textit{Lower panel:} Diffusion coefficient for the transport by shear instability for the same model.
}
\label{fig-coeff6}
\end{figure}

Figures \ref{fig-hrbl} and \ref{fig-stbl} show that the effects of rotation along the birthline are quite small.
This is indeed expected for the following two reasons:
\begin{itemize}
\item The rotation velocities considered here are always far from critical values
and do not imply strong deformations of the accreting star (see Fig.~\ref{fig-vvcm}).
\item Transport processes (meridional currents and shear instabilities) are inefficient during the pre-MS phase
for stars with masses on the ZAMS in the mass range considered here.
Figure \ref{fig-coeff6} illustrates this fact:
it shows the vertical component of the meridional circulation velocity, $U$, and the diffusion coefficient for the transport by shear instability, $D_{\rm shear}$,
for the 6 M$_\odot$ model with $\Omega_{\rm ini}=9\times10^{-6}$ s$^{-1}$ at an age of 200 000 years (just after the end of accretion).
We see that the typical value for $U$ is 10$^{-4}$-10$^{-3}$ cm s$^{-1}$.
For a radius $R\gtrsim5R_\odot$ (this inequality holds for the whole accretion phase),
this leads to a typical timescale for meridional currents equal to $R/U\gtrsim10^7$ yr,
which is much longer than the pre-MS timescale of a few 10$^5$ yr for the 6 M$_\odot$ star.
The timescale for the transport by shear instability is given by $R^2/D_{\rm shear}$.
Figure \ref{fig-coeff6} shows that the typical value for $D_{\rm shear}$ is $10^7-10^8$ cm$^2$ s$^{-1}$, leading to a similar timescale than for meridional circulation.
Therefore, there is no chance for these processes to have a strong impact on the distribution of $\Omega$ and of the chemical elements.
For stars with lower $M_{\rm ZAMS}$, things may be different.
The transport processes have an impact on the surface abundances in light elements (\citealt{eggenberger2012}).
\end{itemize}

\begin{figure}
\includegraphics[width=0.49\textwidth]{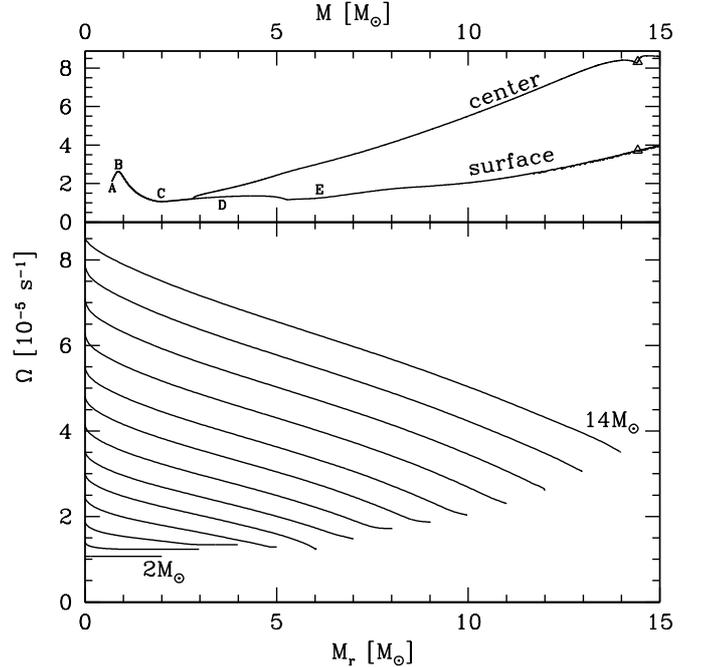}
\caption{
Angular velocity on the birthline with $\rm\Omega_{ini}$ = 2.1$\times$10$^{-5}$ s$^{-1}$ (value
needed to obtain $\rm v_{ZAMS}/v_{crit}=0.40$ for the 14 M$_\odot$).
\textit{Upper panel:} Angular velocity at the surface and in the centre of the star, along the birthline,
as a function of the total mass.
Letters indicate the evolutionary stages described in the text (Sect. \ref{sec-bl}),
and triangles indicate the appearance of the convective core.
\textit{Lower panel:} Rotation profiles on the birthline, at different masses,
from 2 M$_\odot$ to 14 M$_\odot$, by steps of 1 M$_\odot$.
}
\label{fig-ombl}
\end{figure}

Figure \ref{fig-ombl} shows how the angular velocity profile builds up inside the accreting object.
We see that up to a mass of about 3 M$_\odot$, the accreting star rotates as a solid body.
This comes from the fact that the star is completely convective up to this point (see Fig. \ref{fig-stbl}) and we assume that convective zones rotate as solid bodies.
As soon as the radiative core appears and expands, above about 5 M$_\odot$, 
an internal gradient of $\Omega$ develops with a typical ratio of 2 to 3 between the rotation rate of the core and that of the surface.
Note that starting accretion from a core with a lower rotation velocity, produces less steep gradients of $\Omega$ at the end of the accretion phase.
This can be seen comparing the distribution of $\Omega$ in the 6 M$_\odot$ model at the end of the accretion phase when $\Omega_{\rm ini}=9.00\times10^{-6}$ s$^{-1}$
(see Fig.~\ref{fig-om6}) and the curve corresponding to 6 M$_\odot$ in Fig.~\ref{fig-ombl} obtained with $\Omega_{\rm ini}=2.10\times10^{-5}$ s$^{-1}$.

In more general terms, the effect of varying $\rm\Omega_{ini}$ on the value of $\Omega$ during the evolution is the following:
the angular velocity of a given layer at a given evolutionary stage depends almost linearly on $\rm\Omega_{ini}$.
This is due to the fact that, at each time step, the angular velocity of the accreted layer is defined from the previous surface velocity.
Since the transport of angular momentum is negligible on pre-MS timescales, as mentioned above,
the specific angular momentum ($r^2 \Omega$) remains constant and thus
the angular velocity of each layer keeps at each stage a linear dependence on the initial value.

At the end of the accretion phase and for the masses considered here, the star still needs to contract towards the ZAMS.
In the next subsection we describe in more details the complete pre-MS and MS evolution of our models with $M\rm_{ZAMS}$ = 6 and 14 M$_\odot$.

\subsection{Pre-MS of a 6 and 14 M$_\odot$ model with accretion and rotation}
\label{sec-m}

\begin{figure}
\includegraphics[width=0.49\textwidth]{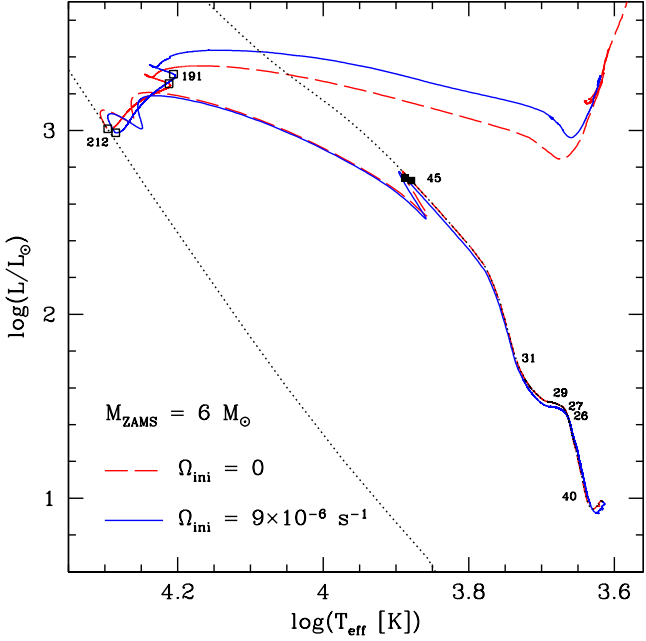}
\caption{
Evolutionary tracks for the models with $M_{\rm ZAMS}$ = 6 M$_\odot$, with rotation (solid blue line) and without rotation (dashed red line).
The numbers along the track of the rotating model indicate the rotation velocity (in km s$^{-1}$)
at M = 1, 2, ..., 6 [M$_\odot$], on the ZAMS, and on the terminal-age main sequence (TAMS).
The right dotted line represents the birthline and the left dotted line the ZAMS of \cite{ekstroem2012} without rotation.
The end of accretion is indicated by filled squares.
The ZAMS and the TAMS of the models are indicated by empty squares.
}
\label{fig-hr6}
\end{figure}

Complete evolutionary tracks for models with $M\rm_{ZAMS}$ = 6 M$_\odot$ computed with and without rotation are shown in Fig. \ref{fig-hr6}.
The initial velocity of the rotating model ($\rm\Omega_{ini}=9\times10^{-6}\,s^{-1}$) is chosen in order to obtain a star rotating at 40\% of the critical velocity on the ZAMS.
This value of $\rm v/v_{crit}=0.4$ on the ZAMS is adopted in the continuity of the computation of a large grid of rotating models at solar metallicity by \cite{ekstroem2012}
and is representative of the rotation velocity of young B-type stars according to \cite{huang2010}.

\begin{figure}
\includegraphics[width=0.49\textwidth]{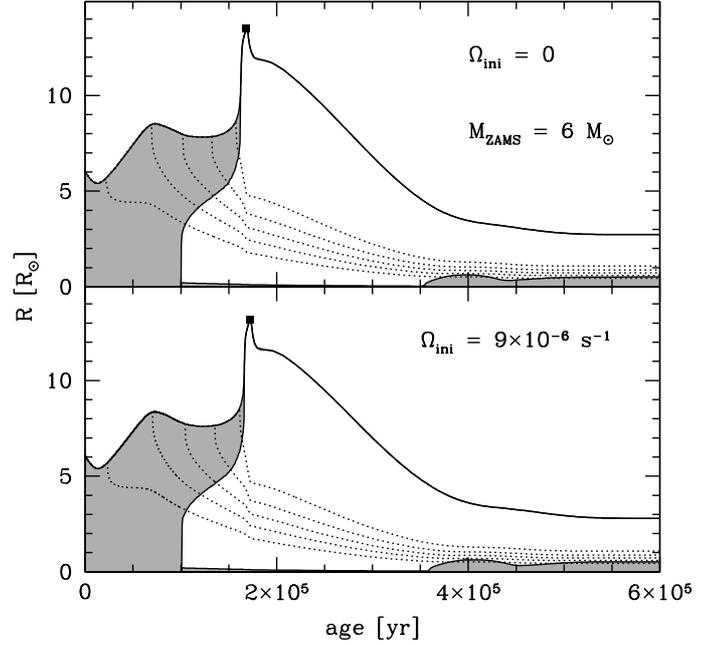}
\caption{Similar figure as Fig. \ref{fig-stbl} following the evolution with time
of the 6 M$_\odot$ model. Time 0 is taken at the beginning of the computation starting
from the 0.7 M$_\odot$ core.
Filled squares indicate the end of accretion.
}
\label{fig-st6}
\end{figure}

The evolution of the radius and of the internal structure of the star during the pre-MS is shown on Fig. \ref{fig-st6},
for the  non-rotating model (upper panel) and the rotating one (lower panel).
The model with $M\rm_{ZAMS}$ = 6 M$_\odot$ follows the birthline
almost until the end of the expansion due to the development of the radiative core (point $E$ on Fig. \ref{fig-stbl}).

We see that the rotating model reaches a given evolutionary stage at a slightly larger age than the non-rotating one.
The age of the star on the ZAMS (counted here from the beginning of the accretion on the 0.7 M$_\odot$ core) is 7.4\% larger with rotation than without rotation.
This effect comes from the fact that the centrifugal force decreases the effective gravity inside the star and thus slows down the contraction.

\begin{figure}
\includegraphics[width=0.49\textwidth]{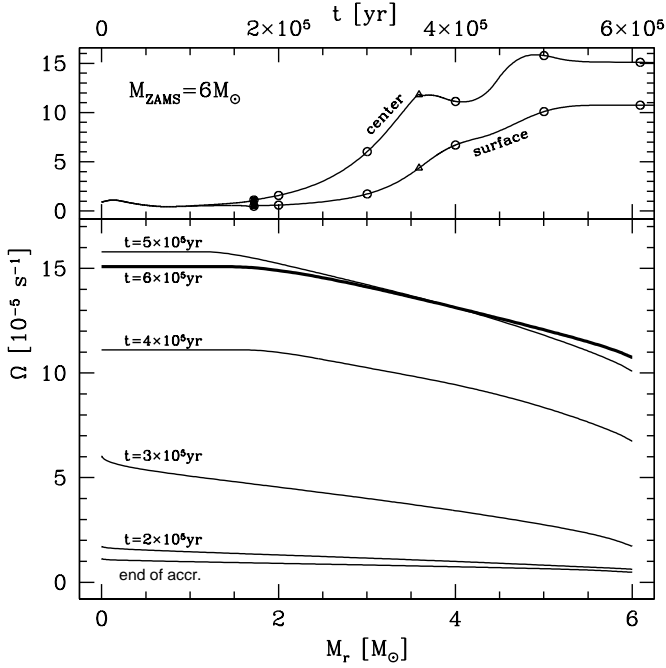}
\caption{
Angular velocity of the model with $M\rm_{ZAMS}$ = 6 M$_\odot$.
\textit{Upper panel:} Angular velocity at the surface and in the centre of the star as a function of the age.
Circles indicate the stages for which the rotation profiles are shown in the lower panel.
The filled square corresponds to the end of accretion, and the triangle indicates the appearance of the convective core.
\textit{Lower panel:} Rotation profiles during the contraction phase,
at different stages indicated by circles on the upper panel. The thickest line corresponds to the ZAMS.
}
\label{fig-om6}
\end{figure}

Fig. \ref{fig-om6} shows the variation of $\Omega$ inside the 6 M$_\odot$ at different stages during the contraction phase.
An interesting question is to know which kind of distribution of $\Omega$ is obtained on the ZAMS.
Is the star rotating as a solid body or not?
When accretion stops, we have a nearly solid-body rotating star.
We see that the contraction phase increases the contrast between the rotation rate of the core and the rotation rate of the envelope.
However, this contrast remains quite modest and the rotation law of the star on the ZAMS is not far from a solid-body rotation law.
The core is rotating 36\% faster than the surface.

\begin{figure}
\includegraphics[width=0.49\textwidth]{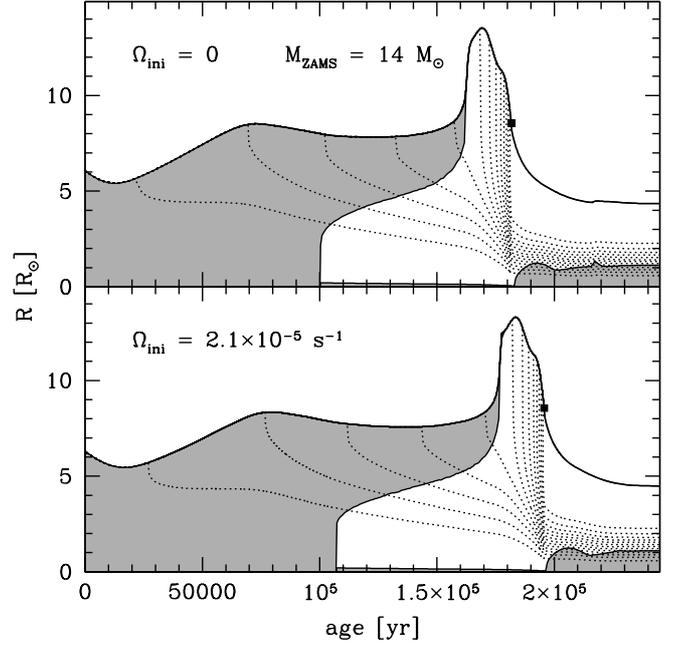}
\caption{Same as Fig. \ref{fig-st6}, but for $M\rm_{ZAMS}$ = 14 M$_\odot$.}
\label{fig-st14}
\end{figure}

The evolution of the structure of the 14 M$_\odot$ model is shown in Fig. \ref{fig-st14}.
Again, the value of $\rm\Omega_{ini}$ is chosen so that $\rm v/v_{crit}=0.4$ on the ZAMS: $\rm\Omega_{ini}=2.1\times10^{-5}\,s^{-1}$.
As in the case of the 6 M$_\odot$ model, we note that a given evolutionary stage
is reached at slightly larger ages when rotation is included.
The effect is however quite small; the time spent on the pre-MS increases by only $3.1\%$ when rotational effects are taken into account.

\begin{figure}
\includegraphics[width=0.49\textwidth]{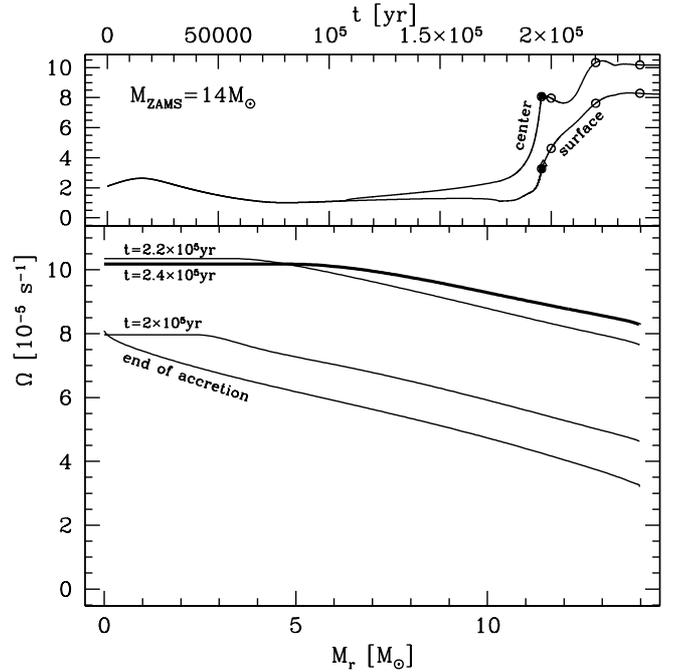}
\caption{Same as Fig. \ref{fig-om6}, but for $M\rm_{ZAMS}$ = 14 M$_\odot$.}
\label{fig-om14}
\end{figure}

Figure \ref{fig-om14} shows the variation of $\Omega$ inside the 14 M$_\odot$ model at different stages during the contraction phase.
In contrast with the results
obtained for the 6 M$_\odot$ model, the contraction phase, which is much shorter than for the
6 M$_\odot$ stellar model, decreases the contrast between the rotation rate of the core and the surface.
Indeed, we see that the ratio of the angular velocity at the center and at the surface is approximately equal to 2 at the end of the accretion, 
while it is about 1.2 on the ZAMS. Thus, in the case of the 14 M$_\odot$ model, the core is rotating about 20\% faster than the surface on the ZAMS,
while it is rotating 36\% faster in the case of the 6 M$_\odot$ model.
For the 2 M$_\odot$ model, again with $\rm v/v_{crit}=0.4$ on the ZAMS (not shown here), the same ratio is about 80\%.
We conclude that in this scenario the degree of differential rotation on the ZAMS decreases when $M\rm_{ZAMS}$ increases
for a given $\rm v/v_{crit}$=0.4 on the ZAMS.

This can be understood by looking at Fig.~\ref{fig-hrbl}. We see that the degree of global contraction, starting from the birthline to the ZAMS,
increases when the mass decreases. This favors stronger gradient of $\Omega$ inside lower mass stars.

\subsection{Grid of rotating pre-MS models}
\label{sec-gr}

\begin{figure}
\includegraphics[width=0.49\textwidth]{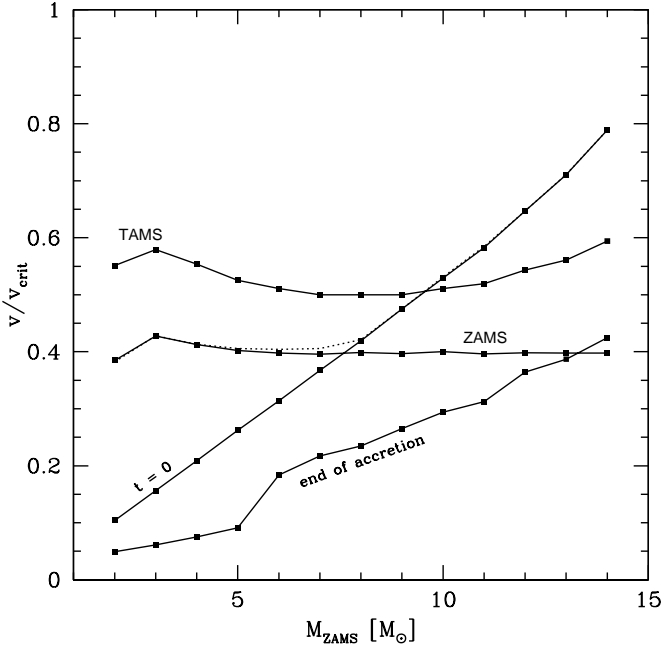}
\caption{
Ratio $\rm v/v_{crit}$ as a function of $M\rm_{ZAMS}$ at different stages of the evolution, for the models described in Sect. \ref{sec-gr}.
The dotted line indicates the maximum value of $\rm v/v_{crit}$ reached during the pre-MS.
}
\label{fig-vvcm}
\end{figure}

We compute the accretion and contraction phases for models with masses $M\rm_{ZAMS}$ between 2 and 14 M$_\odot$, by steps of 1 M$_\odot$.
As explained above, we chose for each final mass a value of $\Omega$ for the initial core such that
an equatorial velocity of about 40\% of the critical velocity is reached on the ZAMS.
These initial values are given in Table \ref{tab-omini}.
\begin{table}
\begin{tabular}{|cc|cc|}
\hline
$M\rm_{ZAMS}$ [M$_\odot$]&$\rm\Omega_{ini}$ [s$^{-1}$]&$M\rm_{ZAMS}$ [M$_\odot$]&$\rm\Omega_{ini}$ [s$^{-1}$]\\
\hline
2&3.00$\times10^{-6}$&  9&1.35$\times10^{-5}$\\
3&4.50$\times10^{-6}$&10&1.50$\times10^{-5}$\\
4&6.00$\times10^{-6}$&11&1.65$\times10^{-5}$\\
5&7.50$\times10^{-6}$&12&1.80$\times10^{-5}$\\
6&9.00$\times10^{-6}$&13&1.95$\times10^{-5}$\\
7&1.05$\times10^{-5}$&14&2.10$\times10^{-5}$\\
8&1.20$\times10^{-5}$&&\\
\hline
\end{tabular}
\caption{Values of $\rm\Omega_{ini}$ for models with different $M\rm_{ZAMS}$ described in Sect. \ref{sec-gr}.}
\label{tab-omini}
\end{table}
The models with $M\rm_{ZAMS}$ of 6 and 14 M$_\odot$ are the same as those described in Sect. \ref{sec-m}.
For the accretion law considered here, the value of $\Omega\rm_{ini}$ increases linearly with the final mass.
The ratio of the surface velocity to the critical velocity at a given time is always below 80\% during the whole pre-MS and MS phase. 
Figure \ref{fig-vvcm} shows the ratio $\rm v/v_{crit}$ as a function of $M\rm_{ZAMS}$ at different stages of the evolution.
There are many interesting points that can be noted from this figure:
\begin{itemize}
\item We see that $\rm v/v_{crit}$ decreases during the accretion process.
This comes from the fact that during most of the accretion phase,
the radius of the accreting star remains around 10 R$_\odot$, while the mass is growing.
Consequently, the surface gravity and hence $\rm v_{crit}$ increase.
\item For low- and intermediate-mass stars ($\rm<8M_{\odot}$), the surface velocity at the end of the accretion phase is still far from the one reached on the ZAMS,
while the difference between these velocities decreases when more massive stars are considered. 
This is quite expected because, for the higher mass range the formation timescale ($M/\dot M$)
becomes of the same order of magnitude as the Kelvin-Helmholtz timescale (typical timescale for the contraction)
and thus the birthline is very close or even follows the ZAMS (\citealt{beech1994}, \citealt{bernasconi1996}).
\item An important point is that such models have difficulties to form stars more massive than about 20 M$_\odot$
and having a surface velocity on the ZAMS equal or superior to about 40\% of the critical velocity.
We see indeed that to obtain such models one should start from initial cores rotating near the critical limit, which does not seem reasonable.
\item A similar problem exists for producing very fast rotators on the ZAMS in the whole mass range considered here.
We develop further this point in Sect.~\ref{sec-vmax}.
\end{itemize}

\begin{figure}
\includegraphics[width=0.49\textwidth]{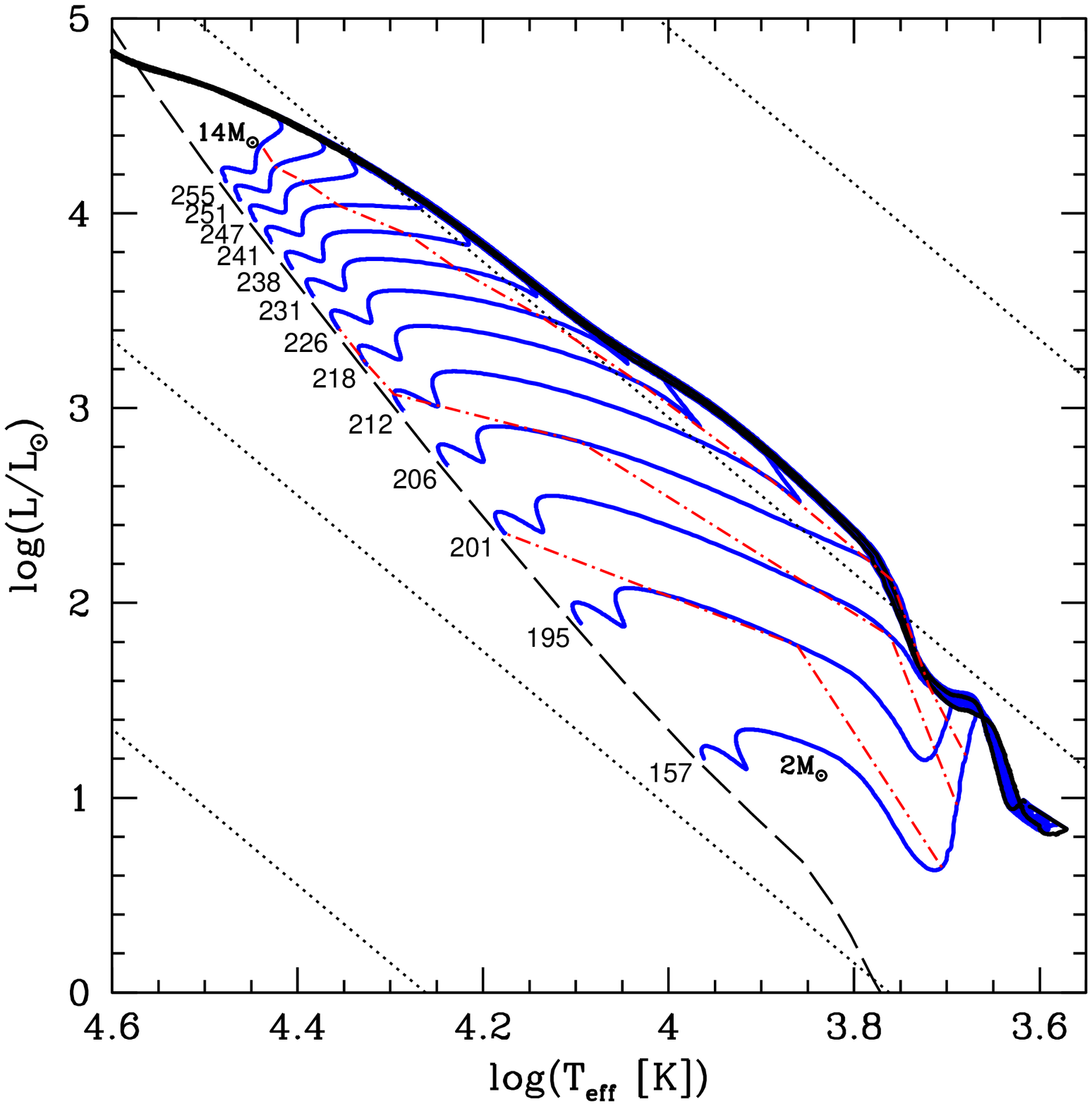}
\caption{
Grid of pre-MS models with $M_{\rm ZAMS}$ from 2 to 14 M$_\odot$, by steps of 1 M$_\odot$,
with $\rm\Omega_{ini}\propto M_{ZAMS}$ (i.e. models described in Sect. \ref{sec-gr}, solid blue lines).
Two birthlines (without rotation and with the higher $\rm\Omega_{ini}$) are plotted (solid black lines).
The dashed line indicates the ZAMS of \cite{ekstroem2012} and dotted lines correspond to iso-radius of 0.1, 1, 10 and 100 R$_\odot$ from left to right.
Equatorial surface velocities on the ZAMS are indicated (in km s$^{-1}$) at the end of each track.
Dot-dashed red lines indicate isochrones of $2\times10^5$, $5\times10^5$ and $2\times10^6$ yr from top to bottom.
}
\label{fig-hrgr}
\end{figure}

The pre-MS tracks of this grid are shown in Fig. \ref{fig-hrgr}.
Tables for the pre-MS evolution of these models can be obtained from the
website [http://obswww.unige.ch/Recherche/evol/-Database-].
An example of a table (the one for $M\rm_{ZAMS}=6\,M_\odot$) is displayed in Table \ref{tab-m6} .

\begin{table*}
\begin{tabular}{|c|c|c|c|c|c|c|c|c|c|c|}
\hline
t [yr]&M [M$_\odot$]&$\rm\dot M$ [M$_\odot$/yr]&R [R$_\odot$]&log$\rm\frac{L}{L_\odot}$&log(T$\rm_{eff}$ [K])&$\rm\Omega_{surf} [s^{-1}]$&$\rm\Omega_{center} [s^{-1}]$&v$\rm_{surf}$ [$\rm\frac{km}{s}$]&$\rm\frac{v}{v_{crit}}$&J [10$^{53}$ g cm$^2$ s$^{-1}$]\\
\hline
$0$                           &$0.7$&$1.38\times10^{-5}$&$6.1$   &$0.98$&$3.62$&$9.03\times10^{-6}$&$9.03\times10^{-6}$&$38$  &$0.31$&$0.0042$\\
$2.350\times10^4$&$1.0$&$1.42\times10^{-5}$&$5.7$  &$1.01$&$3.64$&$9.87\times10^{-6}$&$9.87\times10^{-6}$&$40$  &$0.26$&$0.0059$\\
$7.035\times10^4$&$2.0$&$2.88\times10^{-5}$&$8.3$  &$1.46$&$3.67$&$4.44\times10^{-6}$&$4.44\times10^{-6}$&$26$  &$0.15$&$0.0117$\\
$1.037\times10^5$&$3.0$&$3.06\times10^{-5}$&$7.7$  &$1.50$&$3.69$&$5.04\times10^{-6}$&$5.37\times10^{-6}$&$27$  &$0.12$&$0.0173$\\
$1.349\times10^5$&$4.0$&$3.42\times10^{-5}$&$7.6$  &$1.57$&$3.71$&$5.49\times10^{-6}$&$7.02\times10^{-6}$&$29$  &$0.11$&$0.0229$\\
$1.610\times10^5$&$5.0$&$4.57\times10^{-5}$&$8.5$  &$1.76$&$3.74$&$5.29\times10^{-6}$&$9.08\times10^{-6}$&$31$  &$0.11$&$0.0286$\\
$1.722\times10^5$&$6.0$&$1.89\times10^{-4}$&$13.2$&$2.74$&$3.89$&$4.84\times10^{-6}$&$1.12\times10^{-5}$&$45$  &$0.18$&$0.0345$\\
$1.751\times10^5$&$6.0$&$0$                             &$12.4$&$2.65$&$3.88$&$5.20\times10^{-6}$&$1.16\times10^{-5}$&$45$  &$0.18$&$0.0347$\\
$2.001\times10^5$&$6.0$&$0$                             &$11.5$&$2.65$&$3.89$&$6.00\times10^{-6}$&$1.58\times10^{-5}$&$48$  &$0.18$&$0.0347$\\
$2.279\times10^5$&$6.0$&$0$                             &$10.5$&$2.79$&$3.95$&$7.48\times10^{-6}$&$2.25\times10^{-5}$&$55$  &$0.20$&$0.0347$\\
$2.506\times10^5$&$6.0$&$0$                             &$9.5$  &$2.89$&$4.00$&$9.38\times10^{-6}$&$3.05\times10^{-5}$&$62$  &$0.22$&$0.0347$\\
$2.757\times10^5$&$6.0$&$0$                             &$8.2$  &$3.00$&$4.05$&$1.25\times10^{-5}$&$4.29\times10^{-5}$&$72$  &$0.23$&$0.0347$\\
$3.004\times10^5$&$6.0$&$0$                             &$7.0$  &$3.08$&$4.11$&$1.73\times10^{-5}$&$6.03\times10^{-5}$&$84$  &$0.25$&$0.0347$\\
$3.272\times10^5$&$6.0$&$0$                             &$5.7$  &$3.15$&$4.17$&$2.61\times10^{-5}$&$8.58\times10^{-5}$&$105$&$0.28$&$0.0347$\\
$3.504\times10^5$&$6.0$&$0$                             &$4.8$  &$3.19$&$4.22$&$3.83\times10^{-5}$&$1.11\times10^{-4}$&$130$&$0.32$&$0.0347$\\
$3.750\times10^5$&$6.0$&$0$                             &$4.1$  &$3.17$&$4.25$&$5.43\times10^{-5}$&$1.17\times10^{-4}$&$156$&$0.35$&$0.0347$\\
$4.002\times10^5$&$6.0$&$0$                             &$3.6$  &$3.07$&$4.25$&$6.70\times10^{-5}$&$1.11\times10^{-4}$&$170$&$0.36$&$0.0347$\\
$4.266\times10^5$&$6.0$&$0$                             &$3.4$  &$3.01$&$4.25$&$7.42\times10^{-5}$&$1.15\times10^{-4}$&$177$&$0.37$&$0.0347$\\
$4.502\times10^5$&$6.0$&$0$                             &$3.2$  &$3.05$&$4.27$&$8.14\times10^{-5}$&$1.36\times10^{-4}$&$187$&$0.38$&$0.0347$\\
$4.761\times10^5$&$6.0$&$0$                             &$3.1$  &$3.09$&$4.29$&$9.23\times10^{-5}$&$1.56\times10^{-4}$&$201$&$0.40$&$0.0347$\\
$5.003\times10^5$&$6.0$&$0$                             &$2.9$  &$3.07$&$4.30$&$1.01\times10^{-4}$&$1.58\times10^{-4}$&$210$&$0.40$&$0.0347$\\
$5.251\times10^5$&$6.0$&$0$                             &$2.8$  &$3.02$&$4.29$&$1.06\times10^{-4}$&$1.53\times10^{-4}$&$213$&$0.40$&$0.0347$\\
$5.507\times10^5$&$6.0$&$0$                             &$2.8$  &$3.00$&$4.29$&$1.07\times10^{-4}$&$1.52\times10^{-4}$&$213$&$0.40$&$0.0347$\\
$5.814\times10^5$&$6.0$&$0$                             &$2.8$  &$2.99$&$4.29$&$1.08\times10^{-4}$&$1.51\times10^{-4}$&$213$&$0.40$&$0.0347$\\
$6.091\times10^5$&$6.0$&$0$                             &$2.8$  &$2.99$&$4.29$&$1.08\times10^{-4}$&$1.51\times10^{-4}$&$213$&$0.40$&$0.0347$\\
$9.852\times10^5$&$6.0$&$0$                             &$2.8$  &$2.99$&$4.29$&$1.07\times10^{-4}$&$1.48\times10^{-4}$&$212$&$0.40$&$0.0347$\\
\hline
\end{tabular}
\caption{Table for the model with $M\rm_{ZAMS}$ = 6 M$_\odot$ described in section \ref{sec-gr}.}
\label{tab-m6}
\end{table*}

\begin{figure}
\includegraphics[width=0.49\textwidth]{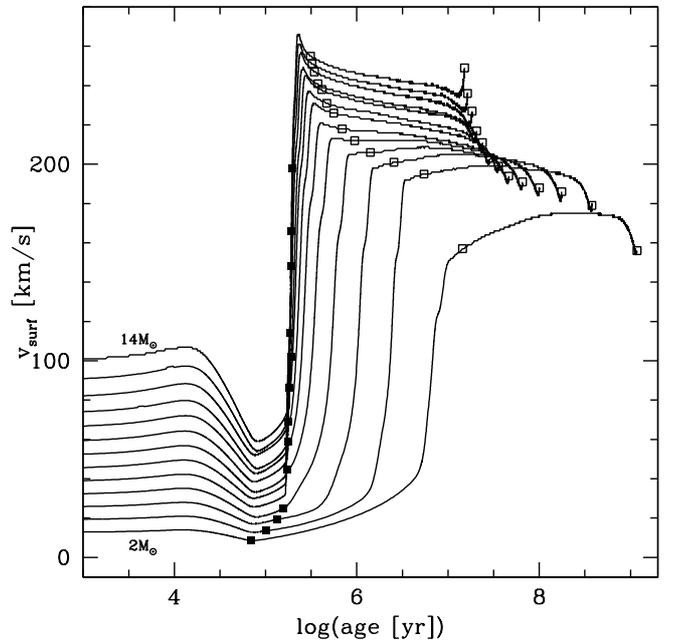}
\caption{
Evolution of the surface velocity of the models described in Sect. \ref{sec-gr}.
Filled squares correspond to the end of the accretion, while empty squares indicate the ZAMS and the TAMS.
}
\label{fig-vgr}
\end{figure}

\begin{figure}
\includegraphics[width=0.49\textwidth]{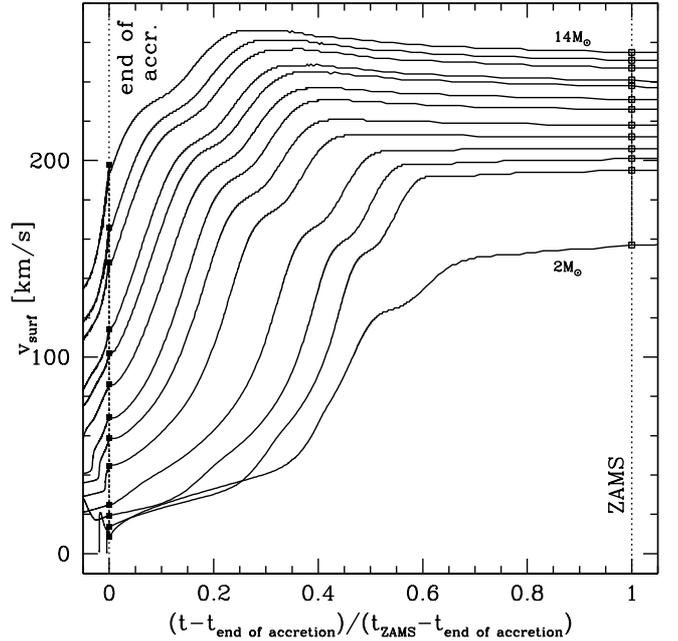}
\caption{
Same as Fig. \ref{fig-vgr}, but focused on the contraction phase.
This time the x-axis is the fraction of the total contraction time,
so that the end of accretion corresponds to 0 and the ZAMS to 1.
}
\label{fig-vgr2}
\end{figure}

The evolution of v$\rm_{surf}$ up to the end of the MS is plotted on Figs. \ref{fig-vgr} and \ref{fig-vgr2} for all these models.
Surface equatorial velocities during the MS phase are typically between 150 and 250 km s$^{-1}$.
These values encompass the observed average equatorial velocities measured for these types of stars on the MS band (\citealt{huang2010}).
This means that the pre-MS evolutions obtained here, provided our accretion history is reasonable,
may be considered as representative for a significant fraction of the stars in this mass range.

We see that these models keep a surface velocity between 10 and 100 km s$^{-1}$ during most of the accretion phase.
A decrease of the surface velocity occurs when the star inflates during central D-burning, because $\dot M$ is still low.
When $\dot M$ becomes high enough, the surface velocity increases, even during the expansion resulting from the growing of the radiative core.
For the low- and intermediate-mass stars, the contraction phase produces a very substantial increase of the surface velocity.
Fig. \ref{fig-vgr2} shows that for the 2 M$_\odot$ stellar model, the surface velocity increases by a factor 8.
For the 14 M$_\odot$, due to the proximity of the
birthline with the ZAMS, the increase is more modest, by only a factor 1.3.
Interestingly, we see that for models with $M_{\rm ZAMS}$ inferior or equal to 7 M$_\odot$,
the surface velocity does not go through a local maximum just before reaching the ZAMS, as it is the case for more massive stars. 
This is related to the fact that the degree of differential rotation on the ZAMS
increases when the mass on the ZAMS decreases.
This reduces the velocity of the meridional currents, which slow down the outer layers and accelerate the inner regions at the very beginning of the MS phase
(see the discussion in Denissenkov et al. 1999).

\subsection{Maximum surface velocities on the ZAMS}
\label{sec-vmax}

As mentioned in Sect. \ref{sec-gr}, in the context of the accretion scenario and with the assumptions considered here,
we cannot form stars with masses $\rm\gtrsim20\,M_\odot$ and $\rm v/v_{crit}\gtrsim0.4$ on the ZAMS.
Even for masses around 10-15 $\rm M_\odot$, there are difficulties to produce fast rotators on the ZAMS.
This is due to the fact that, in this scenario, the ratio $\rm v/v_{crit}$ has two peaks during the pre-MS:
one at the beginning of the birthline and the other one around the ZAMS.
The first peak is responsible for the limitation of the surface velocities on the ZAMS.

\begin{figure}
\includegraphics[width=0.49\textwidth]{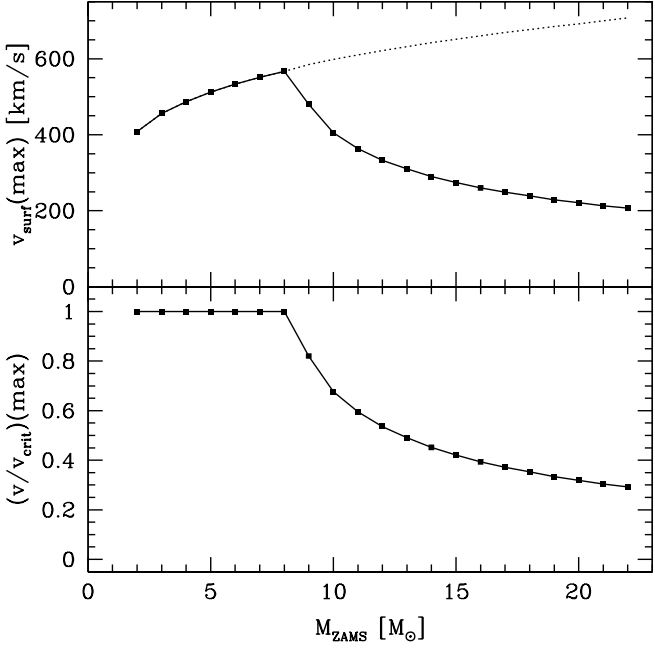}
\caption{
Maximum surface velocities on the ZAMS as a function of $M\rm_{ZAMS}$
for models between 2 and 22 $\rm M_\odot$.
The dotted line in the upper panel indicates the critical velocity.
}
\label{fig-vmaxz}
\end{figure}

For each final mass, we estimate the maximum value for the surface velocity that can be reached on the ZAMS, $\rm v_{surf}(max)$.
The results are shown in Fig. \ref{fig-vmaxz}.
We see that below 8 M$_\odot$, the present scenario does not limit the range of $\rm v/v_{crit}$ that can be obtained on the ZAMS.
For 9 M$_\odot$, the maximum surface velocity on the ZAMS is 82\% of the critical velocity,
while this ratio is 68\% for 10 M$_\odot$.
For such masses, the present scenario cannot produce very fast rotators on the ZAMS.
Above 15 M$_\odot$, the maximum possible value of $\rm v_{ZAMS}/v_{crit}$ is below 0.4.
For a 22 M$_\odot$, the limit is 0.29, which corresponds to $\rm v_{surf}(max)$ = 207 km s$^{-1}$. If evolved on the MS,
such a star would have a time averaged velocity around 150 km s$^{-1}$, below
the typical surface velocity for a star of such a mass (\citealt{huang2010}).

\begin{figure}
\includegraphics[width=0.49\textwidth]{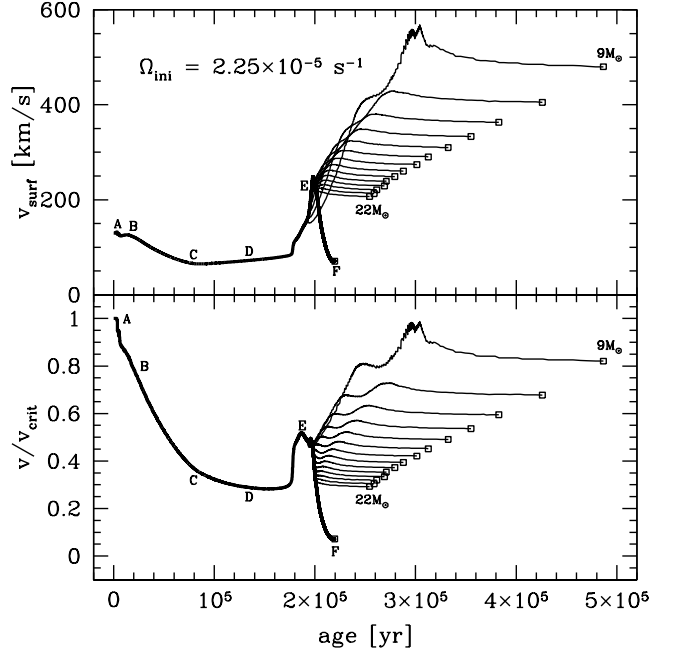}
\caption{
Evolution of the surface velocity during the pre-MS for models with $\rm\Omega_{ini}=2.25\times10^{-5}\,s^{-1}$
and $\rm M_{ZAMS}$ between 9 and 22 $\rm M_\odot$.
The thickest line is the corresponding birthline, and the empty squares indicate the ZAMS.
Letters correspond to the stages described in Sect. \ref{sec-bl}.
}
\label{fig-vmaxpms}
\end{figure}

To make more explicit the cause of this limitation,
we show in Fig. \ref{fig-vmaxpms} how the surface velocity evolves with time
for models ending with different masses on the ZAMS starting with a core rotating at its maximal velocity $\rm\Omega_{ini}=2.25\times10^{-5}\,s^{-1}$.
Such models correspond to the fastest rotators that can be produced with the present accretion scenario.
For $M\rm_{ZAMS}$ $\leq$ 8 M$_\odot$, the critical velocity is reached during the KH contraction phase, thus for these models,
as already mentioned above, it suffices to chose a smaller value of $\rm\Omega_{ini}$ to reach the critical velocity on the ZAMS.
Above 8 $\rm M_\odot$, the star reaches the ZAMS with $\rm v_{surf}(max)<v_{crit}$,
so that velocities between $\rm v_{surf}(max)$ and $\rm v_{crit}$ are not allowed.

We see also that for a given $\rm\Omega_{ini}$ the ZAMS velocity is a decreasing function of $M\rm_{ZAMS}$.
This is related to the fact that the mean specific angular momentum accreted at each time is almost constant along the birthline.
This results in a specific angular momentum at the surface of the star on the ZAMS nearly independent of the mass.
As the radius of a ZAMS star increases with its mass,
the relation $j_{\rm surf}\propto R\times\rm v_{surf}$ leads to a $\rm v_{surf}$ decreasing with the mass on the ZAMS.
Since $\rm v_{crit}$ on the ZAMS increases with the mass, $\rm v/v_{crit}$ on the ZAMS also decreases with the mass.
This is the cause of the problem for fast rotators.
To produce stars with masses superior to 15 M$_\odot$ with $\rm v/v_{crit}\geq0.4$ on the ZAMS,
we need to change the pre-MS scenario.
At that point we leave the problem open.
It is however interesting to note two points:
1) Any disc locking mechanism during the contraction phase will make the problem still more severe;
2) The impossibility to obtain fast rotators concern masses above 8 M$_\odot$.
Does this indicate that more massive stars follow a different accretion history than low- and intermediate-mass stars?
We shall investigate that question in a forthcoming study.
In particular, the effect of varying the accretion law for mass and angular momentum
will be studied in a forthcoming paper (Haemmerl\'{e} et al. in prep.).
We shall see whether a different accretion history will allow to produce fast rotators among massive stars.

Another issue is the impact of the properties of the initial model on the velocity limitation on the ZAMS.
In the present models, all the computations are started with a model corresponding to spherical accretion.
One can wonder if the choice of an initial model built also by disc accretion
could allow to reach higher surface velocities on the ZAMS for massive stars.
Disc accretion allows the accreted material to radiate its internal energy in the disc, before being accreted,
so that a model built by disc accretion has less internal energy, and consequently a smaller radius,
than a model of the same mass built by spherical accretion.
As a consequence, the critical velocity of an initial model built by disc accretion is higher than for the present initial model.
We then can start the computations with a higher $\rm\Omega_{ini}$, but not with a higher angular momentum:
for solid-body rotation, at the critical velocity, one has
\begin{equation}
J=I\times\Omega_{crit}\propto R^2\times R^{-3/2}=R^{1/2}
\end{equation}
so that a smaller radius leads to a smaller angular momentum.
This remark concerns the total angular momentum of the star itself, but also the specific angular momentum of the accreted material,
which remains constant during all the accretion phase.
At the end of the accretion phase, the total amount of angular momentum gathered by a star built from an initial core rotating at its critical velocity
is even smaller if the initial model is built by disc accretion than with the present initial model.
As the radius and the density profile on the ZAMS are fixed,
the surface velocity of a ZAMS star decreases when its angular momentum decreases.
The conclusion is that the velocity limitation on the ZAMS
is even stronger with an initial model coming from disc accretion than with the present one.
It has to be noted that a longer initial phase of spherical accretion would lead to higher radii for a given mass.
Is it possible to significantly relax the velocity limitation on the ZAMS by delaying the transition between spherical and disc accretion phases?
We shall discuss this point in a forthcoming paper.

\section{Comparisons with observed velocities of Herbig Ae/Be stars}
\label{sec-HAeBe}

\begin{figure}
\includegraphics[width=0.49\textwidth]{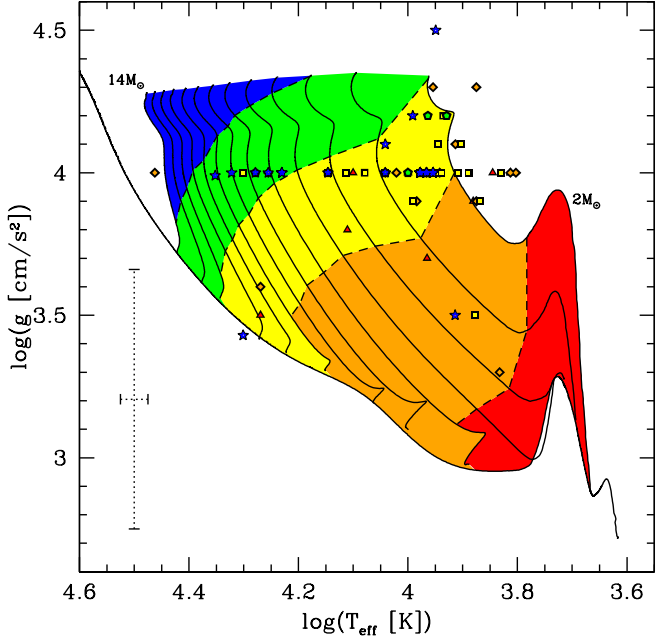}
\caption{
Diagram log T$\rm_{eff}$ - log g.
Solid lines are evolutionary tracks of the grid described in Sect. \ref{sec-gr} (with the non-rotating birthline).
Dashed lines are iso-velocities of 50, 100, 150 and 200 km s$^{-1}$ of these models (from right to left).
Dots indicate observations of Herbig Ae/Be stars with measured v$\cdot$sini from \cite{alecian2012}.
The symbol and its color depends on the value of w=v$\cdot$sini$\cdot$4/$\pi$:
red triangles are used for w $<$ 50 km s$^{-1}$,
orange diamonds for 50 km s$^{-1}$ $<$ w $<$ 100 km s$^{-1}$,
yellow squares for 100 km s$^{-1}$ $<$ w $<$ 150 km s$^{-1}$,
green pentagons for 150 km s$^{-1}$ $<$ w $<$ 200 km s$^{-1}$ and
blue stars for 200 km s$^{-1}$ $<$ w $<$ 250 km s$^{-1}$.
A mean error-bar is displayed, which corresponds to the averaged error over the stars for which an error is given.
}
\label{fig-obs}
\end{figure}

Fig. \ref{fig-obs} shows the pre-MS evolutionary tracks in the Log g versus Log T$\rm_{eff}$ diagram.
In this figure, the bottom line is the birthline and the upper ends of the tracks correspond to the ZAMS.
Theoretically, one expects that pre-MS stars with masses between 2 and 14 M$_\odot$ should be observed in the colored region of the diagram.
The colored regions correspond to different ranges of surface velocities:  
\begin{itemize}
\item red: v $<$ 50 km s$^{-1}$;
\item orange: 50 km s$^{-1}$ $\leq$ v $<$ 100 km s$^{-1}$;
\item yellow: 100 km s$^{-1}$ $\leq$ v $<$ 150 km s$^{-1}$;
\item green: 150 km s$^{-1}$ $\leq$ v $<$ 200 km s$^{-1}$;
\item blue: v $\geq$ 200 km s$^{-1}$.
\end{itemize}
As expected, the surface velocity predicted by the models increases when the surface gravity increases
and when the effective temperature increases.

Overplotted on this diagram are the observed positions of Herbig Ae/Be stars studied by \cite{alecian2012} and \cite{folsom2012}. 
Many stars show a Log $g$ equal to 4.0, which reflects the difficulty of measuring this quantity.
This is also reflected by the large error bars on the determined gravity.
This has of course to be kept in mind when comparisons are made with models.
We see also that the density of points is the greatest in regions corresponding to the lower mass range
where the pre-MS evolution lasts the longer and where the initial mass function predicts a high number of objects. 

To compare the measured rotational velocities with the results of our models, we use different symbols and colors for the observations.
If the predictions of the models were in agreement with these observations, the points of a given color should appear in the region of the corresponding color.
Unsurprisingly, we see that very few observational points lie in their expected region.
Indeed, the track for each mass corresponds to only one peculiar initial condition,
while in reality one expects a distribution of initial conditions, in particular of $\Omega_{\rm ini}$.
Moreover, what is measured is $\rm v\sin i$ and not $\rm v$.
Even if we try to correct for this effect by multiplying the observed $\rm v\sin i$ values by the inverse of the average of $\sin i$ ($4/\pi$),
we cannot pretend that the plotted values correspond to the true velocities. 
So, in the present state, it is not possible to draw any strong conclusion from such comparisons.
We can however note that there appears to be a general trend for having the fastest rotators (blue and green points) on average at a higher gravity and effective temperature
than slow rotators (orange and yellow points), in accordance with the trend expected from the models.
This trend needs however to be confirmed with respect to the difficulties mentioned above.
We hope however that this attempt of comparing the models with the observations will stimulate further observations and theoretical modeling.

\section{Impact of pre-MS evolution on the MS evolution}
\label{sec-MS}

We have seen above that with the accretion law considered here,
the stars do not rotate as solid-body when arriving on the ZAMS, as assumed for example in \cite{ekstroem2012}.
Let us now study the consequence of this small differential rotation in the present models on the evolution during the whole MS phase.
For this purpose, we compute additional models for 6 and 14 M$_\odot$, started on the ZAMS with an initially flat rotation profile (labeled ZAMS), as in \cite{ekstroem2012}.
The models labeled ZAMS(v$\rm_{surf}$) were computed with an $\rm\Omega_{ini}$ chosen such that
v$\rm_{surf}$ on the ZAMS is the same as in the corresponding model with accretion labeled PMS(accr).
In a similar manner, the models denoted ZAMS($J$) were computed with an $\rm\Omega_{ini}$ such that the total angular momentum content, $J$, on the ZAMS,
is the same as in the corresponding model with accretion.

\begin{figure}
\includegraphics[width=0.49\textwidth]{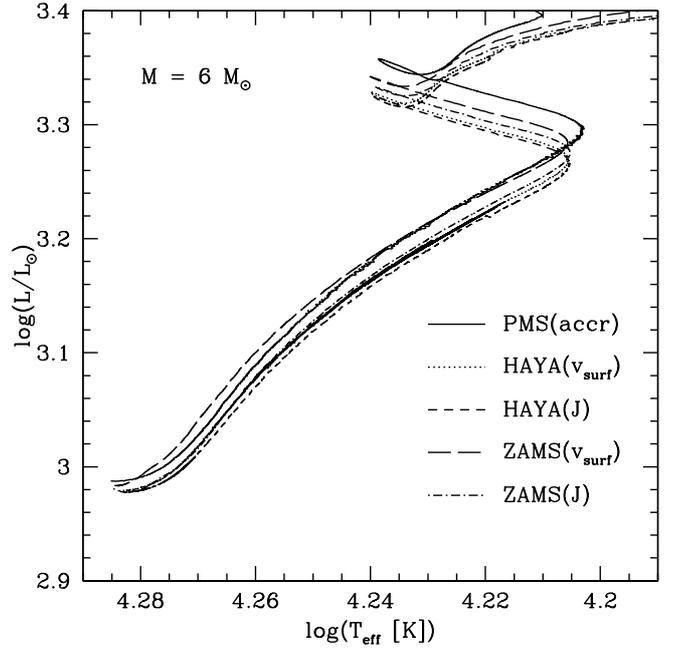}
\caption{HR diagram showing the main-sequence evolution of the five models of 6 M$_\odot$ described in Sect. \ref{sec-MS}.}
\label{fig-hr6sp}
\end{figure}

\begin{figure}
\includegraphics[width=0.49\textwidth]{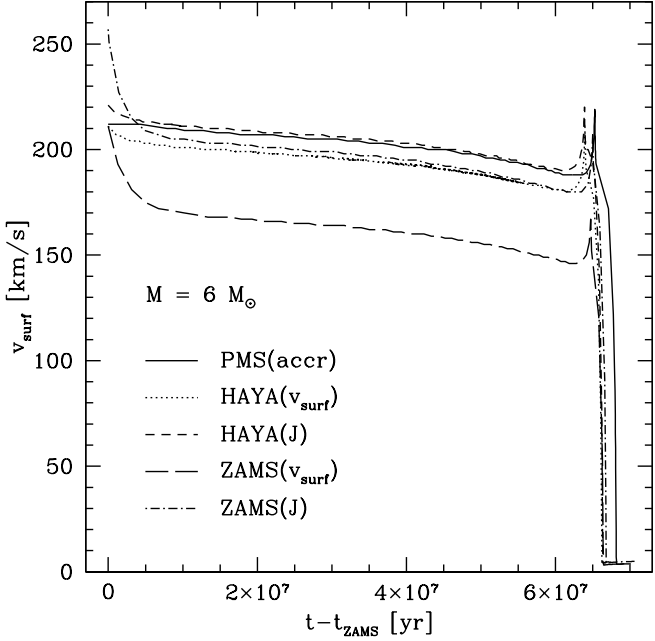}
\caption{Evolution of the surface velocity during the main sequence for the five models of 6 M$_\odot$ described in Sect. \ref{sec-MS}.}
\label{fig-v6sp}
\end{figure}

\begin{figure}
\includegraphics[width=0.49\textwidth]{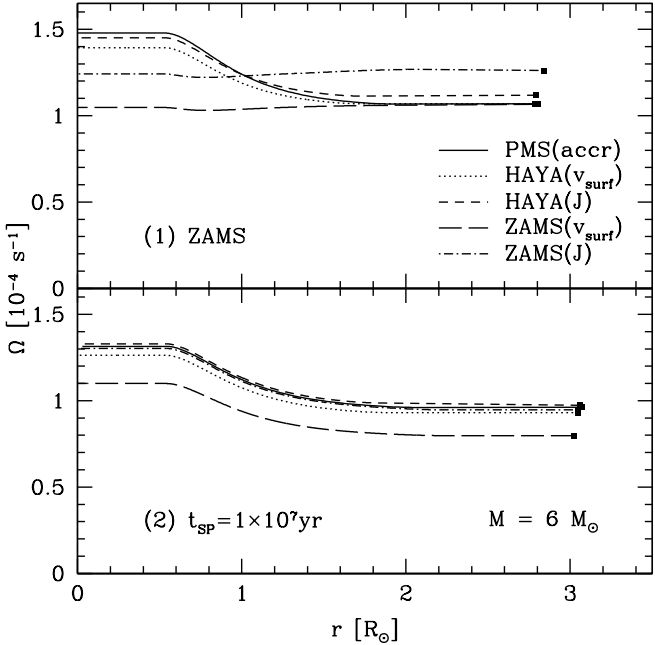}
\caption{
Rotation profiles on the main sequence for the five models of 6 M$_\odot$ described in Sect. \ref{sec-MS}.
The \textit{upper panel} shows the profiles on the ZAMS,
while the \textit{lower panel} shows the profiles 10$^7$ yr after the ZAMS
(i.e. at around 15\% of the time spent on the MS).
Filled squares indicate the surface of the star.
}
\label{fig-om6sp}
\end{figure}

The MS tracks of these different models are displayed on Fig. \ref{fig-hr6sp} for the 6 M$_\odot$ case (with the models labeled HAYA, described below).
A comparison of the evolution of the surface velocity of the same models are shown in Fig. \ref{fig-v6sp}.
Finally, Fig. \ref{fig-om6sp} shows the comparison of their internal rotation profiles.

Fig. \ref{fig-hr6sp} shows that there is no significant difference in the MS tracks of the various models considered.
Comparing the evolution of the surface velocities (see Fig. \ref{fig-v6sp}),
we see that the ZAMS(v$\rm_{surf}$) model presents a strong decrease of the surface velocity at the beginning of the MS phase.
This is due to the fact that when $\Omega$ is constant inside the star,
the meridional currents have a strong velocity and mainly advect angular momentum from the surface to the central region.
This occurs until an equilibrium situation is achieved between this advection process which builds and maintains the gradient of $\Omega$
and the shear turbulence, which tends to erode it (\citealt{zahn1992}, \citealt{denissenkov1999}, \citealt{meynet2000}).

In the present models with accretion and internal transport of angular momentum during the pre-MS phase,
we obtain a shallow internal gradient of $\Omega$ on the ZAMS.
This implies that meridional currents will have an easier job for building the equilibrium profile;
a less strong decrease of the surface velocity at the beginning of the MS phase is then obtained for these models than for models starting from the ZAMS.
This is well apparent in Fig. \ref{fig-v6sp}.
We see that models accounting for the pre-MS phase do not show the decrease of the surface velocity shown by models starting from the ZAMS.

We saw that the gradient of $\Omega$ on the ZAMS in pre-MS models with accretion decreases with the mass. This is why,
for higher masses, a small decrease of the surface velocity can be seen (see the small peak
of velocity just before the ZAMS in Fig.~\ref{fig-vgr}).

The same behavior is found for the ZAMS($J$) model, the main difference being the level of the surface velocity reached when the equilibrium situation is reached.
While in the ZAMS(v$\rm_{surf}$) model, the surface velocity reached after equilibrium is below the one reached in the PMS(accr) model by about 15-20\%,
the surface velocity of the ZAMS($J$) model reaches the same level as in the PMS(accr) model.

Importantly, the comparison of the evolution of the surface velocities between the models ZAMS($\rm v_{surf}$) and PMS(accr)
shows that two stars with the same initial mass, position in the HR diagram, and initial surface velocity may 
show very different evolutions for the surface velocities during the MS phase depending on their total content of angular momentum.
When the total angular momentum is the same in both models, the evolution of the surface velocities are the same except at the very beginning of the MS phase.

It is interesting now to look at the evolution of the internal distribution of $\Omega$ and of the surface chemical enrichments.
These two features are strongly related since
the main drivers of the surface enrichments are the gradients of $\Omega$. 
Comparing the distributions of $\Omega$, we note the following differences:
\begin{itemize}
\item After 15\% of the total MS time, the rotation profiles of all the models are similar in terms of $\nabla\Omega$.
\item At this stage, the only remaining differences are simple shifts in the absolute value of $\Omega$,
due to the differences of total angular momentum content of the star.
\item While this shift is negligible in ZAMS($J$) models,
it is more pronounced in ZAMS(v$\rm_{surf}$) models ($\sim$15-20\%).
\end{itemize}
This shift in the rotation profile is at the origin of the shift in v$\rm_{surf}$ previously mentioned.
Except for this global shift, we see that the various rotation profiles converge in a time short compared to the total MS time.
As a consequence, there is no significant differences in the surface enrichments during the MS phase of all these models.
These remarks are true for both the 6 and 14 M$_\odot$ models.

From the above considerations, we can conclude that the assumption of solid-body rotation on the ZAMS is justified
and provides evolution during the MS phase quite in agreement with the evolution obtained from models accounting for the pre-MS phase with accretion.
However, this is true only if models to be compared start with the same total angular momentum content on the ZAMS.

It is interesting also to compare the PMS(accr) models with models starting on the Hayashi line and evolving at constant mass on the ZAMS.
Such initial conditions would mimic stars following a very strong initial episode of accretion, which would bring the model at the top of the Hayashi line.
In a similar manner to what was done above, we computed models labeled HAYA(v$\rm_{surf}$) and HAYA($J$), starting on the Hayashi line with
initial rotation rates such that v$\rm_{surf}$ on the ZAMS, respectively $J$ on the ZAMS, are the same as in PMS(accr) models.
Results of these models are shown in Figs. \ref{fig-hr6sp}, \ref{fig-v6sp} and \ref{fig-om6sp}.
We see that the results obtained from the HAYA models are all quite similar to the results of the 
PMS(accr) model. In contrast with the ZAMS models, they do not show a very marked decrease of the surface velocity
at the beginning of the MS phase as already noted above. This comes from the fact that stars are not rotating as solid bodies at that stage (see Fig. \ref{fig-om6sp}).
A noticeable difference with respect to the PMS(accr) models is that the HAYA models allow to reach much higher velocities on the ZAMS. This will
be studied in a forthcoming work with accretion rates very strong in a first phase and then decreasing with time.

\section{Conclusions and perspectives}
\label{sec-Outro}

We have studied the pre-MS evolution of stars with masses on the ZAMS between 2 and 22 M$_\odot$
taking into account rotation and accretion in the frame of the cold disc accretion scenario, with an accretion rate increasing with the mass.
Our main results are the following:
\begin{itemize}
\item Stars reach the ZAMS with a rotation profile that is not far from solid-body rotation,
with small (but non-zero) gradient of angular velocity.
The core rotates between 20\% (14 M$_\odot$) and 80\% (2 M$_\odot$) faster than the surface for a fixed ratio $\rm v/v_{crit}$=0.4 on the ZAMS.
\item For all the cases considered in the present work, models of a given mass and total angular momentum content
present very similar MS evolutions in terms of evolutionary tracks, surface velocities and abundances, whatever their formation history.
This is fortunate since if it would not be the case, then the evolution of stars, in addition to depend on the mass, metallicity, rotation,
magnetic field, multiplicity would also depend on the formation history. This does not appear to be the case.
\item Above 8 M$_\odot$, the present scenario cannot produce the whole range of surface velocities on the ZAMS. Only velocities
below an upper limit can be achieved. For instance, it is not possible to produce 15 M$_\odot$ with rotational velocities on the ZAMS equal
or higher than $\rm v/v_{crit}$=0.4.
\end{itemize}
This last conclusion leads to the question of the origin of the massive fast rotators.
We can identify different lines of research for future studies:
1) As was briefly indicated above, it may be that a different history for the mass accretion rate may be a solution;
2) Also here we supposed that the accreted layer is deposited onto the star with the same angular momentum as the surface.
In the disc, matter rotates with a Keplerian velocity, which is equivalent to the critical velocity.
It may be that the matter accreted have a specific angular momentum higher than the surface.
This would of course change the results obtained here even keeping the same mass accretion law;
3) Finally, it may be also that some stars acquire their high velocity through tidal interactions or mass transfer in close binaries (\citealt{demink2012}).

\acknowledgements{Part of this work was supported by the Swiss National Science Foundation.}

\bibliographystyle{aa}

\bibliography{bibliotheque}

\begin{thebibliography}{36}
\expandafter\ifx\csname natexlab\endcsname\relax\def\natexlab#1{#1}\fi

\bibitem[{{Alecian} {et~al.}(2012){Alecian}, {Wade}, {Catala}, {Grunhut},
  {Landstreet}, {Bagnulo}, {B{\"o}hm}, {Folsom}, {Marsden}, \&
  {Waite}}]{alecian2012}
{Alecian}, E., {Wade}, G.~A., {Catala}, C., {et~al.} 2012, \mnras, 394

\bibitem[{{Asplund} {et~al.}(2005){Asplund}, {Grevesse}, \&
  {Sauval}}]{asplund2005}
{Asplund}, M., {Grevesse}, N., \& {Sauval}, A.~J. 2005, in Astronomical Society
  of the Pacific Conference Series, Vol. 336, Cosmic Abundances as Records of
  Stellar Evolution and Nucleosynthesis, ed. T.~G. {Barnes}, III \& F.~N.
  {Bash}, 25

\bibitem[{{Beech} \& {Mitalas}(1994)}]{beech1994}
{Beech}, M. \& {Mitalas}, R. 1994, \apjs, 95, 517

\bibitem[{{Behrend} \& {Maeder}(2001)}]{behrend2001}
{Behrend}, R. \& {Maeder}, A. 2001, \aap, 373, 190

\bibitem[{{Bernasconi}(1997)}]{bernasconi1997}
{Bernasconi}, P.~A. 1997, PhD thesis, Universit{\'e} de Gen{\`e}ve

\bibitem[{{Bernasconi} \& {Maeder}(1996)}]{bernasconi1996}
{Bernasconi}, P.~A. \& {Maeder}, A. 1996, \aap, 307, 829

\bibitem[{{Bodenheimer} \& {Sweigart}(1968)}]{bodenheimer1968}
{Bodenheimer}, P. \& {Sweigart}, A. 1968, \apj, 152, 515

\bibitem[{{Caughlan} \& {Fowler}(1988)}]{caughlan1988}
{Caughlan}, G.~R. \& {Fowler}, W.~A. 1988, Atomic Data and Nuclear Data Tables,
  40, 283

\bibitem[{{Churchwell}(1998)}]{churchwell1998}
{Churchwell}, E. 1998, The Origin of Stars and Planetary Systems (ed. C. Lada,
  \& N. Kylafis)

\bibitem[{{Cunha} {et~al.}(2006){Cunha}, {Hubeny}, \& {Lanz}}]{cunha2006}
{Cunha}, K., {Hubeny}, I., \& {Lanz}, T. 2006, \apjl, 647, L143

\bibitem[{{de Mink} {et~al.}(2012){de Mink}, {Langer}, {Izzard}, {Sana}, \& {de
  Koter}}]{demink2012}
{de Mink}, S.~E., {Langer}, N., {Izzard}, R.~G., {Sana}, H., \& {de Koter}, A.
  2012, ArXiv e-prints

\bibitem[{{Denissenkov} {et~al.}(1999){Denissenkov}, {Ivanova}, \&
  {Weiss}}]{denissenkov1999}
{Denissenkov}, P.~A., {Ivanova}, N.~S., \& {Weiss}, A. 1999, \aap, 341, 181

\bibitem[{{Eggenberger} {et~al.}(2012){Eggenberger}, {Haemmerl{\'e}}, {Meynet},
  \& {Maeder}}]{eggenberger2012}
{Eggenberger}, P., {Haemmerl{\'e}}, L., {Meynet}, G., \& {Maeder}, A. 2012,
  \aap, 539, A70

\bibitem[{{Eggenberger} {et~al.}(2008){Eggenberger}, {Meynet}, {Maeder},
  {Hirschi}, {Charbonnel}, {Talon}, \& {Ekstr{\"o}m}}]{eggenberger2008}
{Eggenberger}, P., {Meynet}, G., {Maeder}, A., {et~al.} 2008, \apss, 316, 43

\bibitem[{{Ekstr{\"o}m} {et~al.}(2012){Ekstr{\"o}m}, {Georgy}, {Eggenberger},
  {Meynet}, {Mowlavi}, {Wyttenbach}, {Granada}, {Decressin}, {Hirschi},
  {Frischknecht}, {Charbonnel}, \& {Maeder}}]{ekstroem2012}
{Ekstr{\"o}m}, S., {Georgy}, C., {Eggenberger}, P., {et~al.} 2012, \aap, 537,
  A146

\bibitem[{{Folsom} {et~al.}(2012){Folsom}, {Bagnulo}, {Wade}, {Alecian},
  {Landstreet}, {Marsden}, \& {Waite}}]{folsom2012}
{Folsom}, C.~P., {Bagnulo}, S., {Wade}, G.~A., {et~al.} 2012, \mnras, 422, 2072

\bibitem[{{Girichidis} {et~al.}(2011){Girichidis}, {Federrath}, {Banerjee}, \&
  {Klessen}}]{girichidis2011}
{Girichidis}, P., {Federrath}, C., {Banerjee}, R., \& {Klessen}, R.~S. 2011,
  \mnras, 413, 2741

\bibitem[{{Henning} {et~al.}(2000){Henning}, {Lapinov}, {Schreyer}, {Stecklum},
  \& {Zinchenko}}]{henning2000}
{Henning}, T., {Lapinov}, A., {Schreyer}, K., {Stecklum}, B., \& {Zinchenko},
  I. 2000, \aap, 364, 613

\bibitem[{{Hosokawa} \& {Omukai}(2009)}]{hosokawa2009}
{Hosokawa}, T. \& {Omukai}, K. 2009, \apj, 691, 823

\bibitem[{{Hosokawa} {et~al.}(2010){Hosokawa}, {Yorke}, \&
  {Omukai}}]{hosokawa2010}
{Hosokawa}, T., {Yorke}, H.~W., \& {Omukai}, K. 2010, \apj, 721, 478

\bibitem[{{Huang} {et~al.}(2010){Huang}, {Gies}, \& {McSwain}}]{huang2010}
{Huang}, W., {Gies}, D.~R., \& {McSwain}, M.~V. 2010, \apj, 722, 605

\bibitem[{{Kuiper} {et~al.}(2010){Kuiper}, {Klahr}, {Beuther}, \&
  {Henning}}]{kuiper2010}
{Kuiper}, R., {Klahr}, H., {Beuther}, H., \& {Henning}, T. 2010, \apj, 722,
  1556

\bibitem[{{Kuiper} {et~al.}(2011){Kuiper}, {Klahr}, {Beuther}, \&
  {Henning}}]{kuiper2011}
{Kuiper}, R., {Klahr}, H., {Beuther}, H., \& {Henning}, T. 2011, \apj, 732, 20

\bibitem[{{Larson}(1969)}]{larson1969}
{Larson}, R.~B. 1969, \mnras, 145, 271

\bibitem[{{McNally}(1964)}]{mcnally1964}
{McNally}, D. 1964, \apj, 140, 1088

\bibitem[{{Meynet} \& {Maeder}(2000)}]{meynet2000}
{Meynet}, G. \& {Maeder}, A. 2000, \aap, 361, 101

\bibitem[{{Nakano}(1989)}]{nakano1989}
{Nakano}, T. 1989, \apj, 345, 464

\bibitem[{{Peters} {et~al.}(2011){Peters}, {Banerjee}, {Klessen}, \& {Mac
  Low}}]{peters2011}
{Peters}, T., {Banerjee}, R., {Klessen}, R.~S., \& {Mac Low}, M.-M. 2011, \apj,
  729, 72

\bibitem[{{Peters} {et~al.}(2010){Peters}, {Klessen}, {Mac Low}, \&
  {Banerjee}}]{peters2010}
{Peters}, T., {Klessen}, R.~S., {Mac Low}, M.-M., \& {Banerjee}, R. 2010, \apj,
  725, 134

\bibitem[{{Shu}(1977)}]{shu1977}
{Shu}, F.~H. 1977, \apj, 214, 488

\bibitem[{{Stahler} {et~al.}(1986{\natexlab{a}}){Stahler}, {Palla}, \&
  {Salpeter}}]{stahler1986ii}
{Stahler}, S.~W., {Palla}, F., \& {Salpeter}, E.~E. 1986{\natexlab{a}}, \apj,
  308, 697

\bibitem[{{Stahler} {et~al.}(1986{\natexlab{b}}){Stahler}, {Palla}, \&
  {Salpeter}}]{stahler1986i}
{Stahler}, S.~W., {Palla}, F., \& {Salpeter}, E.~E. 1986{\natexlab{b}}, \apj,
  302, 590

\bibitem[{{Stahler} {et~al.}(1980{\natexlab{a}}){Stahler}, {Shu}, \&
  {Taam}}]{stahler1980i}
{Stahler}, S.~W., {Shu}, F.~H., \& {Taam}, R.~E. 1980{\natexlab{a}}, \apj, 241,
  637

\bibitem[{{Stahler} {et~al.}(1980{\natexlab{b}}){Stahler}, {Shu}, \&
  {Taam}}]{stahler1980ii}
{Stahler}, S.~W., {Shu}, F.~H., \& {Taam}, R.~E. 1980{\natexlab{b}}, \apj, 242,
  226

\bibitem[{{Yorke} \& {Bodenheimer}(2008)}]{yorke2008}
{Yorke}, H.~W. \& {Bodenheimer}, P. 2008, in Astronomical Society of the
  Pacific Conference Series, Vol. 387, Massive Star Formation: Observations
  Confront Theory, ed. H.~{Beuther}, H.~{Linz}, \& T.~{Henning}, 189

\bibitem[{{Zahn}(1992)}]{zahn1992}
{Zahn}, J.-P. 1992, \aap, 265, 115

\end{thebibliography}

\end{document}